\documentclass[twocolumn,secnumarabic,amssymb, nobibnotes, aps, pra]{revtex4-1}

\usepackage{amsmath,amssymb,graphicx}
\usepackage{textgreek}

\begin{document}

\title{Analytical study of coherence in seeded modulation instability}

\author{J. Bonetti}
\author{S. M. Hernandez}
\affiliation{Instituto Balseiro, Av. Bustillo km 9.500, Bariloche (R8402AGP), Argentina}

\author{P. I. Fierens}
\affiliation{Instituto Tecnol\'ogico de Buenos Aires, Eduardo Madero 399, Buenos Aires (C1106ACD), Argentina}
\affiliation{Consejo Nacional de Investigaciones Cient\'ificas y T\'ecnicas (CONICET), Argentina}

\author{D. F. Grosz}
\affiliation{Instituto Balseiro, Av. Bustillo km 9.500, Bariloche (R8402AGP), Argentina}
\affiliation{Consejo Nacional de Investigaciones Cient\'ificas y T\'ecnicas (CONICET), Argentina}

\begin{abstract}
We derive analytical expressions for the coherence in the onset of modulation instability, in excellent agreement with thorough numerical simulations. As usual, we start by a linear perturbation analysis, where broadband noise is added to a continuous wave (CW) pump; then, we investigate the effect of adding a deterministic seed to the CW pump, a case of singular interest as it is commonly encountered in parametric amplification schemes. Results for the dependence of coherence on parameters such as fiber type, pump power, propagated distance, seed signal-to-noise ratio are presented. Finally, we show the importance of including higher-order linear and nonlinear dispersion when dealing with generation in longer wavelength regions (mid IR). We believe these results to be of relevance when applied to the analysis of the coherence properties of supercontinua generated from CW pumps.

\end{abstract}

\pacs{42.65,42.81}

\date{\today}%
\maketitle

\section{Introduction}

Supercontinuum (SC) generation has been widely studied in the last few years (see, \textit{e.g.}, \cite{Dudley.RMP.2006,Dudley.SCBook.2010} and references therein). In particular, the relation between modulation instability (MI) and SC has been a matter of profound analysis (see, for example, \cite{Demircan.OpticsComm.2005,Demircan.AppliedPhysB.2007}). Indeed, MI is found to be a source of initial spectral broadening as it leads to the break up of an input pulse into several sub-pulses. Moreover, modulation instability, being seeded by the noise at the input, is one of the sources of fluctuations in supercontinuum spectra. We must note that the relation of pump noise to supercontinuum generation has been widely studied (see, \textit{e.g.}, \cite{Corwin.ApplPhysB.2003,Newbury.OpticsLett.2003,Abeeluck.AppliedPhysicsLetters.2004,Vanholsbeeck.OptExpress.2005,Dudley.OptExpress.2008,Kobtsev.OptExpress.2008,Solli.PhysRevLett.2008,Solli.NaturePhotonics.2012} and references therein).

One relevant aspect of SC spectra is its phase coherence. One metric of phase coherence is given by \cite{Dudley.OpticsLett.2002,Dudley.IEEEJournalSelectedTopicsQuantumElectr.2002}
\begin{equation}
g_{12}(z,\Omega)=  \frac{\left<\tilde{a}_k^*(z,\Omega)\tilde{a}_l(z,\Omega)\right>_{k\neq l}}{\sqrt{\left<\left|\tilde{a}_k(z,\Omega)\right|^2\right>\left<\left|\tilde{a}_l(z,\Omega)\right|^2\right>}},
\label{eq:sccoherence}
\end{equation}
where $\tilde{a}(z,\Omega)$ is the Fourier transform of the pulse envelope at position $z$ along the fiber, the subscripts $k,l$ correspond to different realizations and the angle brackets denote ensemble averages. There is a rich literature on ways of improving this metric. For example, S\o rensen \textit{et al.} \cite{Sorensen.JOSAB.2012} study the effect of seeding the input pump with one or more smaller sideband signals. 

In this work, we study analytically the metric in Eq.~\eqref{eq:sccoherence}. Our development follows the same path as the usual modulation instability analysis, \textit{i.e.}, we investigate a perturbation to a continuous wave. As such, our results are valid only under the undepleted-pump approximation. However, we believe that they provide valuable insight into the coherence of supercontinuum spectra as they allow to characterize the initial fluctuations. 

In Section \ref{sec:analytical}, we develop analytical expressions for $g_{12}$ when the input is a noisy perturbation to a continuous wave. Specifically, the input perturbation spectrum is given by $\tilde{a}(0,\Omega) = \tilde{s}(\Omega)+ \tilde{N}(\Omega)$, where $\tilde{s}(\Omega)$ is the Fourier transform of the deterministic seed and $\tilde{N}(\Omega)$ corresponds to additive white Gaussian noise. Such noise may be considered as an approximation to the shot noise typical of a laser output. Since expressions are somewhat complex, we provide examples of their use for some particular inputs and limiting cases. In Section \ref{subsec:singlesideband}, we specialize the equations to the case in which $\tilde{s}(\Omega) = 0$ for $\Omega < 0$. This situation corresponds to the very important case of a seed wavelength alongside with the pump. Section \ref{subsec:equalpowersidebands} presents results corresponding the case of symmetrical power spectrum, \textit{i.e.}, $|\tilde{s}(\Omega)| = |\tilde{s}(-\Omega)|$. 

A comparison of the analytical results with simulations is presented in Section \ref{sec:simulations}. Final conclusions are presented in Section \ref{sec:conclusions}.

\section{Broadband noise as a perturbation to aCW pump}
\label{sec:analytical}

Wave propagation in a lossless optical fibers can be described by the generalized nonlinear Schr\"odinger equation (GNLSE) \cite{Agrawal.NLFO.2012},
\begin{equation}
\frac{\partial A}{\partial z}-i\hat{\beta}A
= i \hat{\gamma} A(z,T)\int\limits_{-\infty}^{+\infty}R(T') \left|A(z,T-T')\right|^2 dT',
\label{eq:gnlse}
\end{equation}
where $A(z,T)$ is the slowly-varying envelope, $z$ is the spatial coordinate, and $T$ is the time coordinate in a comoving frame at the group velocity ($=\beta_1^{-1}$). $\hat{\beta}$ and $\hat{\gamma}$ are operators related to the dispersion and nonlinearity, respectively, and are defined by
\begin{equation*} 
\hat{\beta}= \sum_{m\geq 2}\frac{i^m}{m!}\beta_m \frac{\partial^m }{\partial T^m},\; \hat{\gamma} = \sum_{n \geq 0} \frac{i^n}{n!} \gamma_n \frac{\partial^n }{\partial T^n}.
\end{equation*}
The $\beta_m$'s are the coefficients of the Taylor expansion of the propagation constant $\beta(\omega)$ about a central frequency $\omega_0$. In the convolution integral in the right hand side of Eq.~\eqref{eq:gnlse}, $R(T)$ is the nonlinear response function including both the instantaneous (electronic) and delayed Raman response.

Modulation instability is customarily analyzed by studying the evolution of a small perturbation $a$ to the stationary solution of Eq.~\eqref{eq:gnlse}:
\begin{equation}
A(z,T) = \left(\sqrt{P_0}+a(z,T)\right) e^{i\gamma_0P_0z}.
\label{eq:perturbation}
\end{equation}
By only keeping terms linear in the perturbation, after some straightforward algebra, substitution of Eq.~\eqref{eq:perturbation} into Eq.~\eqref{eq:gnlse} leads to the following 2\textsuperscript{nd} order ordinary differential equation in the frequency domain:
\begin{equation}
\frac{\partial^2 \tilde{a}(z,\Omega)}{\partial z^2} +2i\tilde{B}(\Omega)\frac{\partial \tilde{a}(z,\Omega)}{\partial z}
- \tilde{C}(\Omega) \tilde{a}(z,\Omega) = 0,
\label{eq:gnlseomegaq}
\end{equation}
where $\Omega = \omega - \omega_0$, $\tilde{a}$ and $\tilde{R}$ are the Fourier transforms of $a$ and $R$, respectively, where, for the sake of clarity, we have introduced the variables
\begin{align*}
\tilde{B}(\Omega)=&-\left[\tilde{\beta}_o(\Omega)+P_0 \tilde{\gamma}_o(\Omega)\left(1+\tilde{R}(\Omega)\right)\right],\\
\tilde{C}(\Omega)= &\tilde{\beta}_o^2(\Omega)-\tilde{\beta}_e^2(\Omega)+\\
+&P_0^2\left(\tilde{\gamma}_o^2(\Omega)-\tilde{\gamma}_e^2(\Omega)\right)\left(1+2\tilde{R}(\Omega)\right)-P_0^2\gamma_0^2+\\
+&2P_0\gamma_0 \tilde{\beta}_e(\Omega)+2P_0^2\gamma_0 \tilde{\gamma}_e(\Omega)\left(1+\tilde{R}(\Omega)\right)+\\
+&2P_0\left(\tilde{\beta}_o\tilde{\gamma}_o-\tilde{\beta}_e\tilde{\gamma}_e\right)\left(1+\tilde{R}(\Omega)\right),\\
\tilde{\beta}_e(\Omega) = &\sum_{n\geq 1} \frac{\beta_{2n}}{(2n)!} \Omega^{2n},\;\hat{\beta}_o(\Omega) = \sum_{n\geq 1} \frac{\beta_{2n+1}}{(2n+1)!} \Omega^{2n+1},\\
\tilde{\gamma}_e(\Omega) = &\sum_{n\geq 0} \frac{\gamma_{2n}}{(2n)!} \Omega^{2n},\;\hat{\gamma}_o(\Omega) = \sum_{n\geq 0} \frac{\gamma_{2n+1}}{(2n+1)!} \Omega^{2n+1}.
\end{align*}
Substitution of the ansatz $a(z,\Omega) = D\exp(iK(\Omega)z)$ in Eq.~\eqref{eq:gnlseomegaq} leads to
\begin{equation*}
K^2(\Omega)+2K(\Omega)\tilde{B}(\Omega)+\tilde{C}(\Omega) = 0.
\end{equation*}
Finally, the dispersion relation obtained as a solution to this equation is
\begin{equation}
K(\Omega) = -\tilde{B}(\Omega)\pm \sqrt{\tilde{B}^2(\Omega)-\tilde{C}(\Omega)}.
\label{eq:reldispa}
\end{equation}
This expression agrees with those found in the literature (see, \textit{e.g.}, \cite{Bejot.PhysRevA.2011}). As usual, MI gain can be found as the imaginary part of $K(\Omega)$. 

Some further calculations show that the spectral evolution of the perturbation can be written as 
\begin{equation}
\begin{split}
\tilde{a}(z,\Omega) = &\frac{e^{-i\tilde{B}(\Omega)z}}{K_D(\Omega)}\times\\
&\;\times\left[D_z(\Omega)\tilde{a}^*(0,-\Omega)+F_z(\Omega)\tilde{a}(0,\Omega)\right],
\end{split}
\label{eq:perturbationspect8}
\end{equation}
where $K_D(\Omega) = \sqrt{\tilde{B}^2(\Omega)-\tilde{C}(\Omega)}$ and
\begin{align*}
D_z(\Omega) =&iP_0\tilde{\gamma}(\Omega)\tilde{R}(\Omega) \sin(K_D(\Omega)z),\\
F_z(\Omega) =&i\left(\tilde{\beta}_e+P_0\tilde{\gamma}_e(1+\tilde{R})-P_0\gamma_0\right)\sin(K_Dz)+\\
&+K_D\cos(K_Dz).
\end{align*}

Let us assume that $a(0,\Omega) = \tilde{s}(\Omega)+\tilde{N}(\Omega)$, where $\tilde{s}(\Omega)$ is the Fourier transform of a deterministic seed and $\tilde{N}(\Omega)$ are independent and identically distributed circularly symmetric normal variables with variance $\sigma^2$ for each $\Omega$, \textit{i.e.}, $\tilde{N}(\Omega) \sim \mathcal{CN}(0,\sigma^2)$. If $a_k(z,\Omega)$, $k\in \mathbb{N}$, are independent realizations, then, by Eq.~\eqref{eq:perturbationspect8}
\begin{equation}
\begin{split}
&\left<\tilde{a}_k^*(z,\Omega) \tilde{a}_l(z,\Omega)\right> = \left|\frac{e^{-i\tilde{B}(\Omega)z}}{K_D(\Omega)}\right|^2 \times \\
&\qquad\times \left\{\left|D_z(\Omega)\right|^2 |\tilde{s}(-\Omega)|^2+\left|F_z(\Omega)\right|^2\left|\tilde{s}(\Omega)\right|^2+\right.\\
&\qquad\;\;\left.+2\mathrm{Re}\left\{D_z(\Omega)F_z^*(\Omega)\tilde{s}(\Omega)\tilde{s}(-\Omega)\right\} \right\},
\label{eq:awgn3}
\end{split}
\end{equation}
for $k\neq l$. In the case $k=l$, 
\begin{equation}
\begin{split}
&\left<\left|\tilde{a}_k^2(z,\Omega)\right|^2\right> = \left|\frac{e^{-i\tilde{B}(\Omega)z}}{K_D(\Omega)}\right|^2 \times \\
&\qquad\times\left\{\left|D_z(\Omega)\right|^2 \left(|\tilde{s}(-\Omega)|^2+\sigma^2\right)+\right.\\
&\left.\qquad\;\;+\left|F_z(\Omega)\right|^2\left(\left|\tilde{s}(\Omega)\right|^2+\sigma^2\right)+\right.\\
&\left.\qquad\;\;+2\mathrm{Re}\left\{D_z(\Omega)F_z^*(\Omega)\tilde{s}(\Omega)\tilde{s}(-\Omega)\right\} \right\}.
\end{split}
\label{eq:awgn4}
\end{equation}
Using Eqs.~\eqref{eq:awgn3}-\eqref{eq:awgn4}, the coherence of the perturbation becomes
\begin{equation}
\begin{split}
&\frac{1}{g_{12}(z,\Omega)} = 1 + \left(\left|D_z(\Omega)\right|^2+\left|F_z(\Omega)\right|^2\right)\sigma^2\times\\
&\qquad\times \left\{\left|D_z(\Omega)\right|^2 |\tilde{s}(-\Omega)|^2+\left|F_z(\Omega)\right|^2\left|\tilde{s}(\Omega)\right|^2+\right.\\
&\qquad\;\;\left.+2\mathrm{Re}\left\{D_z(\Omega)F_z^*(\Omega)\tilde{s}(\Omega)\tilde{s}(-\Omega)\right\} \right\}^{-1}.
\end{split}
\label{eq:awgn5}
\end{equation} 
Note that $g_{12}$ depends only on $\beta_{2k}$. 

\subsection{Single sideband}
\label{subsec:singlesideband}

As a relevant example of the use of this equation, consider the case in which, for a given $\Omega$, $\tilde{s}(-\Omega) = 0$ and $\tilde{s}(\Omega) \neq 0$ (\textit{e.g.}, $\tilde{s}(\Omega) = 0$ for $\Omega < 0$). This case is of particular interest because since besides the pump there is a smaller seed signal at a given wavelength, a situation commonly found in various parametric amplification schemes. Then,
\begin{equation}
\frac{1}{g_{12}(z,\Omega)} = 1 +\left[1+\left(\frac{\left|D_z(\Omega)\right|^2}{\left|F_z(\Omega)\right|^2}\right)^{\mathrm{sgn}(\Omega)}\right] \frac{\sigma^2}{\left|\tilde{s}(|\Omega|)\right|^2},
\label{eq:awgn6}
\end{equation} 
where $\mathrm{sgn}(\cdot)$ denotes the sign function. When $|\tilde{s}(|\Omega|)|^2 \gg \sigma^2$,
\begin{equation*}
g_{12}(z,\Omega) \approx 1 -\left[1+\left(\frac{\left|D_z(\Omega)\right|^2}{\left|F_z(\Omega)\right|^2}\right)^{\mathrm{sgn}(\Omega)}\right] \frac{\sigma^2}{\left|\tilde{s}(|\Omega|)\right|^2}.
\label{eq:awgn7}
\end{equation*}

It may be easier to understand these expressions in the particular case where $\gamma_n = 0$ for $n\geq 1$ (\textit{e.g.}, no self-steepening) and $f_R = 0$ (\textit{i.e.}, only electronic instantaneous response). If we further assume that there exists MI gain, 
\begin{equation*}
\begin{split}
&\frac{1}{g_{12}(z,\Omega)} = 1 +\left[1+\right.\\
+&\left.\left(\frac{\gamma_0^2P_0^2\sinh^2(g(\Omega)z/2)}{\gamma_0^2P_0^2\sinh^2(g(\Omega)z/2)+\frac{1}{4}g^2(\Omega)}\right)^{\mathrm{sgn}(\Omega)} \right]\frac{\sigma^2}{\left|\tilde{s}(|\Omega|)\right|^2},
\end{split}
\label{eq:awgn6.1}
\end{equation*} 
where $g(\Omega)=2K_D(\Omega)/i$ is the modulation gain. Interesting expressions are obtained in two limiting cases, when 
either $g(\Omega)z \ll 1$ (propagated distance much shorter than the MI length) or $g(\Omega)z \gg  1$ (propagated distance much longer than the MI length). In the former case,
\begin{equation*}
\begin{split}
\gamma_0^2P_0^2\sinh^2(g(\Omega)z/2) &\approx \gamma_0^2P_0^2(g(\Omega)z/2)^2 \\
&\approx \frac{g^2(\Omega)}{4} \left(\frac{z}{L_{\mathrm{NL}}}\right)^2,
\end{split}
\end{equation*}
where  as $L_{\mathrm{NL}} = (\gamma_0P_0)^{-1}$ is the usual nonlinear length. Thus,
\begin{equation*}
g_{12}(z,\Omega) \approx 1 -\left[1+\left(1+\left(\frac{L_{\mathrm{NL}}}{z}\right)^2\right)^{-\mathrm{sgn}(\Omega)}\right] \frac{\sigma^2}{\left|\tilde{s}(|\Omega|)\right|^2}.
\label{eq:awgn6.4}
\end{equation*} 
Note that coherence does not depend on the actual MI-gain. If we further assume that $z \ll L_{\mathrm{NL}}$,
\begin{equation*}
g_{12}(z,\Omega) \approx 
\begin{cases}
1 -\left(1+\left(\frac{z}{L_{\mathrm{NL}}}\right)^2\right) \frac{\sigma^2}{\left|\tilde{s}(|\Omega|)\right|^2}& \Omega >0,\\
1 -\left(2+\left(\frac{L_{\mathrm{NL}}}{z}\right)^2\right) \frac{\sigma^2}{\left|\tilde{s}(|\Omega|)\right|^2}& \Omega <0.
\end{cases}
\label{eq:awgn6.6}
\end{equation*} 
It is clear from this equation that, while coherence decreases as $z$ increases for $\Omega > 0$, it increases with $z$ for $\Omega < 0$. 

In the case where $g(\Omega)z\gg 1$, coherence depends neither on the particular MI gain nor on the sign of $\Omega$, and 
\begin{equation}
g_{12}(z,\Omega) \approx 1 - 2\frac{\sigma^2}{\left|\tilde{s}(|\Omega|)\right|^2}.
\label{eq:awgn6.3}
\end{equation} 
In this respect, as $z$ increases, coherence of Stokes and anti-Stokes components approach the same value. 

\subsection{Equal power spectrum sidebands}
\label{subsec:equalpowersidebands}

Another particular case is when $|\tilde{s}(-\Omega)| = |\tilde{s}(\Omega)|$. This case corresponds to that where the pump is modulated by two smaller signals of the same amplitude, located symmetrical to the pump (in the frequency space.) In this case,
\begin{equation*}
\begin{split}
\frac{1}{g_{12}(z,\Omega)} = 1 + &\frac{1}{1+\frac{2\mathrm{Re}\left\{D_z(\Omega)F_z^*(\Omega)e^{i(\phi_s(\Omega)+\phi_s(-\Omega))}\right\}}{\left|D_z(\Omega)\right|^2+\left|F_z(\Omega)\right|^2}}\times\\
&\qquad\times\frac{\sigma^2}{\left|\tilde{s}(\Omega)\right|^2},
\end{split}
\label{eq:awgn8}
\end{equation*} 
where $\phi_s(\Omega) = \mathrm{arg}\{\tilde{s}(\Omega)\}$. It is interesting to note that coherence depends on the phase relation: $\phi_s(\Omega)+\phi_s(-\Omega)$. In particular, coherence is maximized when 
\begin{equation*}
\mathrm{Re}\left\{D_z(\Omega)F_z^*(\Omega)e^{i(\phi_s(\Omega)+\phi_s(-\Omega))}\right\} = |D_z(\Omega)||F_z(\Omega)|.
\label{eq:awgn9}
\end{equation*}
So, we can write that
\begin{equation*}
\frac{1}{g_{12}(z,\Omega)} \geq 1 + \frac{1}{1+\frac{2\left|D_z(\Omega)\right|\left|F_z(\Omega)\right|}{\left|D_z(\Omega)\right|^2+\left|F_z(\Omega)\right|^2}}\frac{\sigma^2}{\left|\tilde{s}(\Omega)\right|^2}.
\label{eq:awgn10}
\end{equation*} 
From this expression, it is simple to prove that
\begin{equation}
g_{12}(z,\Omega) \leq \frac{1}{1 + \frac{1}{2}\frac{\sigma^2}{\left|\tilde{s}(\Omega)\right|^2}} \approx 1 - \frac{1}{2}\frac{\sigma^2}{\left|\tilde{s}(\Omega)\right|^2},
\label{eq:awgn12}
\end{equation} 
where the approximation is valid for large input signal-to-noise ratio. 

It is interesting to compare Equations \eqref{eq:awgn6.3} and \eqref{eq:awgn12}. It might be argued that the double-sideband case may lead to a higher spectral coherence than the single-sideband one. However, it must be noted that the higher coherence can be achieved by tuning the phase of the input seed. In the particular case of two input pulses at frequencies symmetrical with respect to the pump, only a particular phase relationship maximizes coherence.

\section{Results}
\label{sec:simulations}

\begin{figure}[ht]
\includegraphics[width=\linewidth]{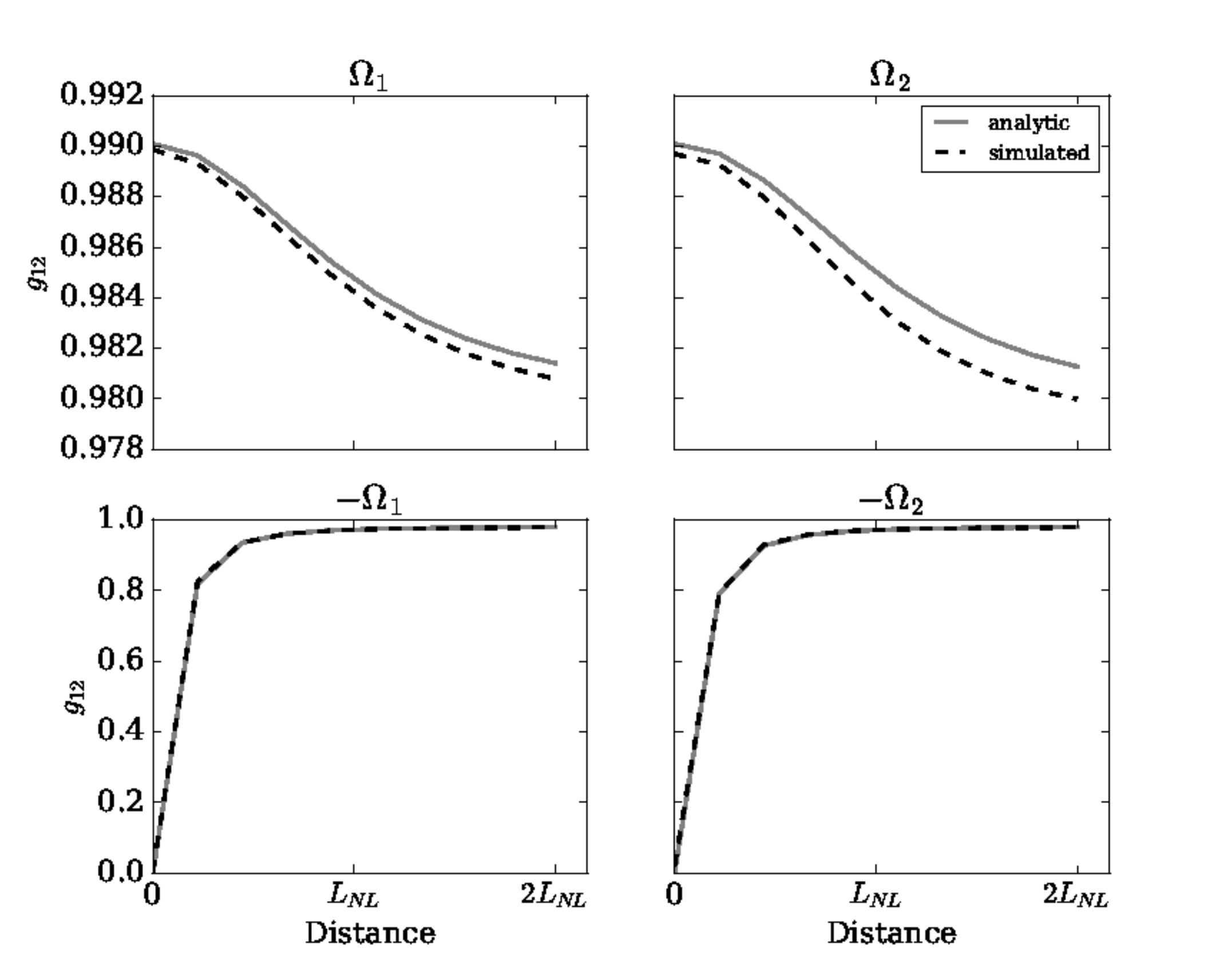}
\caption{$g_{12}(\Omega)$ versus distance. Analytical (full line) and numerical (dashed line) results. Seed frequencies are $\Omega= \pm \Omega_{1,2}$ for $\Omega_1=31$ GHz and $\Omega_2=46$ GHz. Fiber type is SSMF.}
\label{fig:1}
\end{figure}

In order to test the validity of our analytical results, in Fig.~\ref{fig:1} we compare the value of $g_{12}$ by seeding with 1 mW power at frequencies $31$ GHz and $46$ GHz, and show its evolution along a standard Single-Mode Fiber (SSMF, $\beta_2=-21$ ps\textsuperscript{2}/km, zero-dispersion wavelength at 1310 nm, and $\gamma_0 = 1.2$ (W-km)\textsuperscript{-1}) over a distance of  $2 L_{\mathrm{NL}}$ for both the Stokes and anti-Stokes sidebands, and for a pump power of 1 W at 1550 nm. It can be seen that the analytically evaluated $g_{12}$ (Eq.~\eqref{eq:awgn6}) is in excellent agreement with the one calculated from 300 noise realizations (a total of $150 \times 299$ averaged correlations), where the seed signal-to-noise ratio is 10 dB. Also, the `same limit' property stated in Eq.~\eqref{eq:awgn6.3} is verified.

\begin{figure}[hb]
\includegraphics[width=\linewidth]{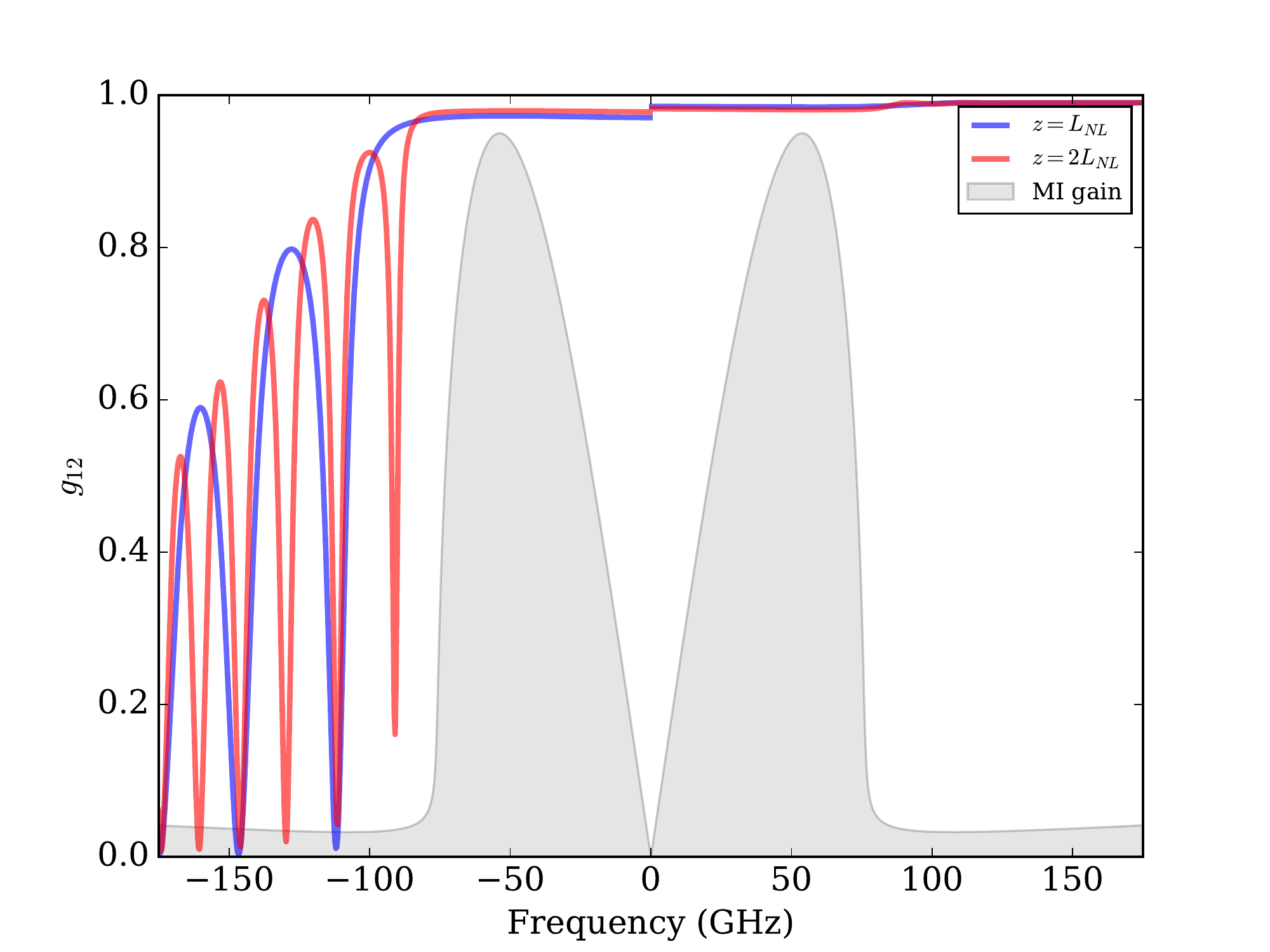}
\caption{$g_{12}(\Omega)$ versus frequency at distances $L_{\mathrm{NL}}$ and $2 L_{\mathrm{NL}}$, and for a SSMF fiber ($P_0=1$ W).}
\label{fig:2}
\end{figure}

\begin{figure}[hb]
\includegraphics[width=\linewidth]{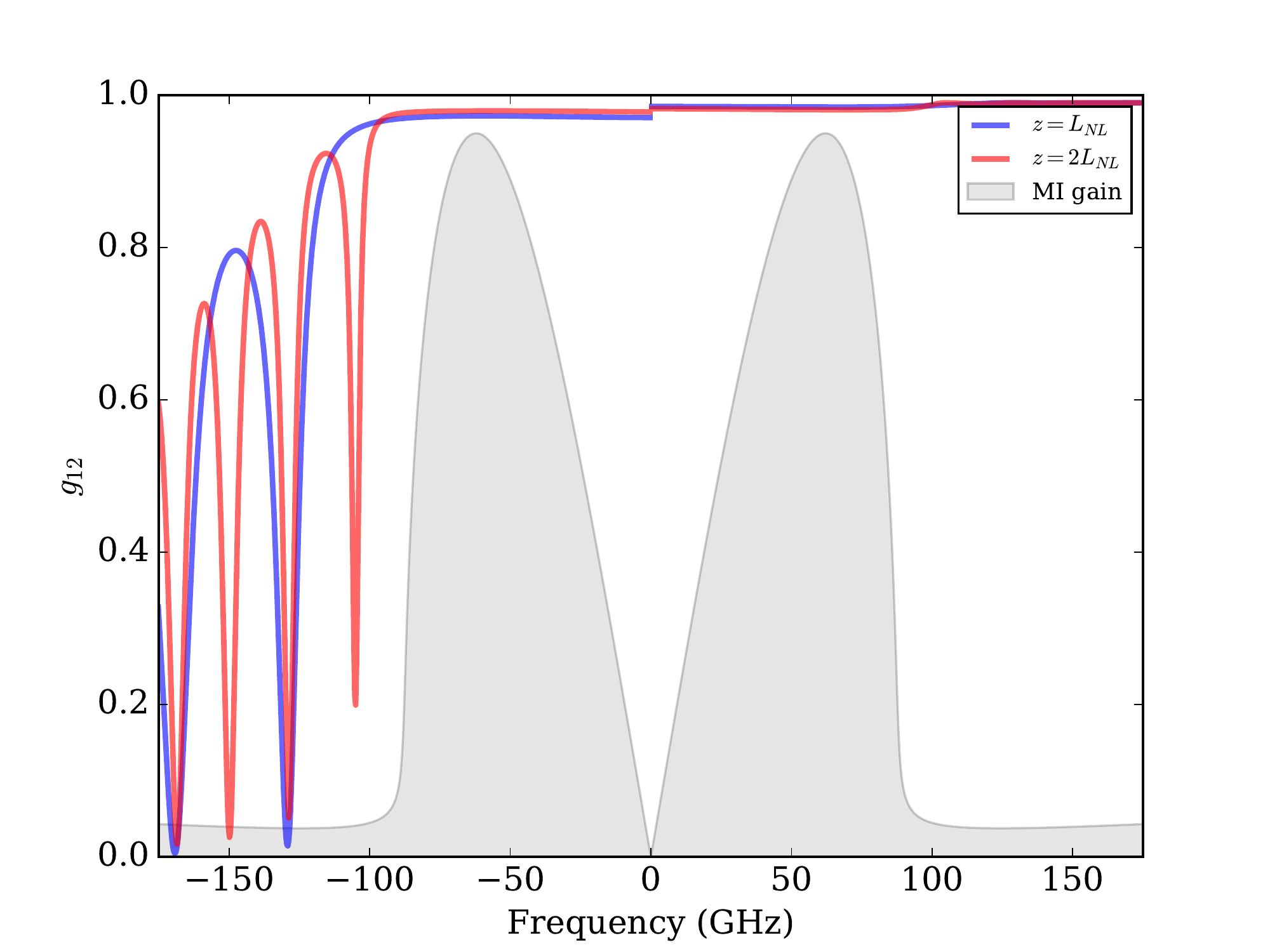}
\caption{$g_{12}(\Omega)$ versus frequency at distances $L_{\mathrm{NL}}$ and $2 L_{\mathrm{NL}}$, and for a NZ DSF fiber ($P_0=1$ W).}
\label{fig:3}
\end{figure}

In Figs.~\ref{fig:2}--\ref{fig:3}, we depict $g_{12}(\Omega)$ when seeding with 1 mW power at frequencies $\Omega > 0$, 10 dB input signal-to-noise ratio, and for distances $L_{\mathrm{NL}}$ and $2L_{\mathrm{NL}}$. While in Fig.~\ref{fig:2} we show results for a SSMF fiber, in Fig.~\ref{fig:3}, for comparison, results shown correspond to a typical non-zero dispersion-shifted fiber (NZ-DSF, same $\beta_2$ as SSMF, zero-dispersion wavelength at 1450 nm, $\gamma_0 = 1.6$ (W-km)\textsuperscript{-1}). By looking at the lower frequency side of the spectrum, we see that coherence (\textit{i.e.}, the value of $g_{12}$) increases with distance up to the limit given by Eq.~\eqref{eq:awgn6.3}, as expected. A `coherence bandwidth', defined as the frequency range from the pump frequency to the first coherence minimum, is observed to dimish with increasing distance for both fiber types, limiting the region where high coherence components are generated. The width of this region is strongly dependent on the MI gain profile (overlayed for comparison), and clearly the higher nonlinearity of the NZ-DSF yields a larger coherent bandwidth as compared to SSMF.

\begin{figure}
\includegraphics[width=\linewidth]{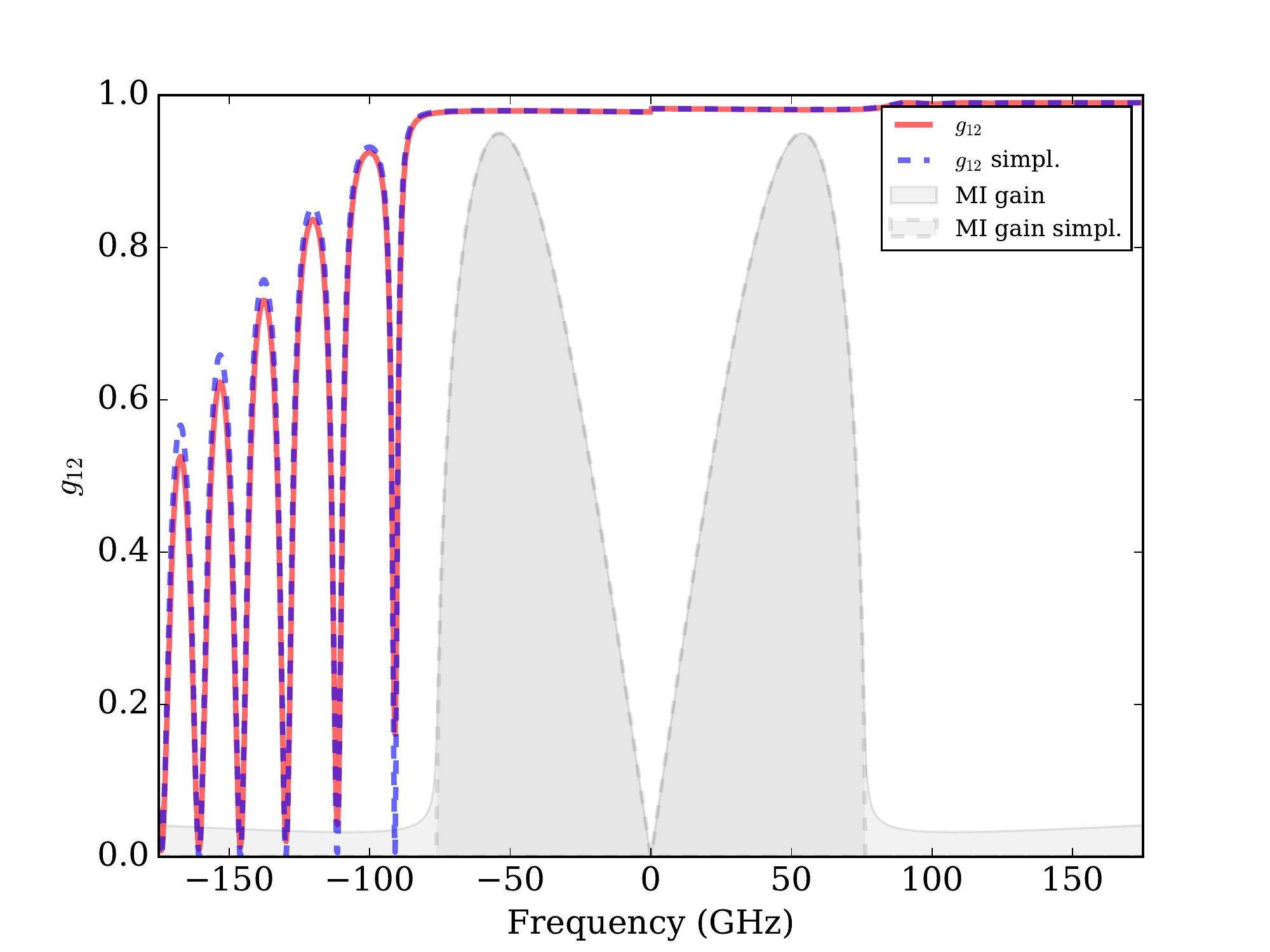}
\caption{$g_{12}(\Omega)$ at distance $2 L_{\mathrm{NL}}$ for a simplified fiber ($\beta_{n\geq 3}=0$, no self-steepening, and no Raman) versus complete fiber model.}
\label{fig:4}
\end{figure}

\begin{figure}
\includegraphics[width=\linewidth]{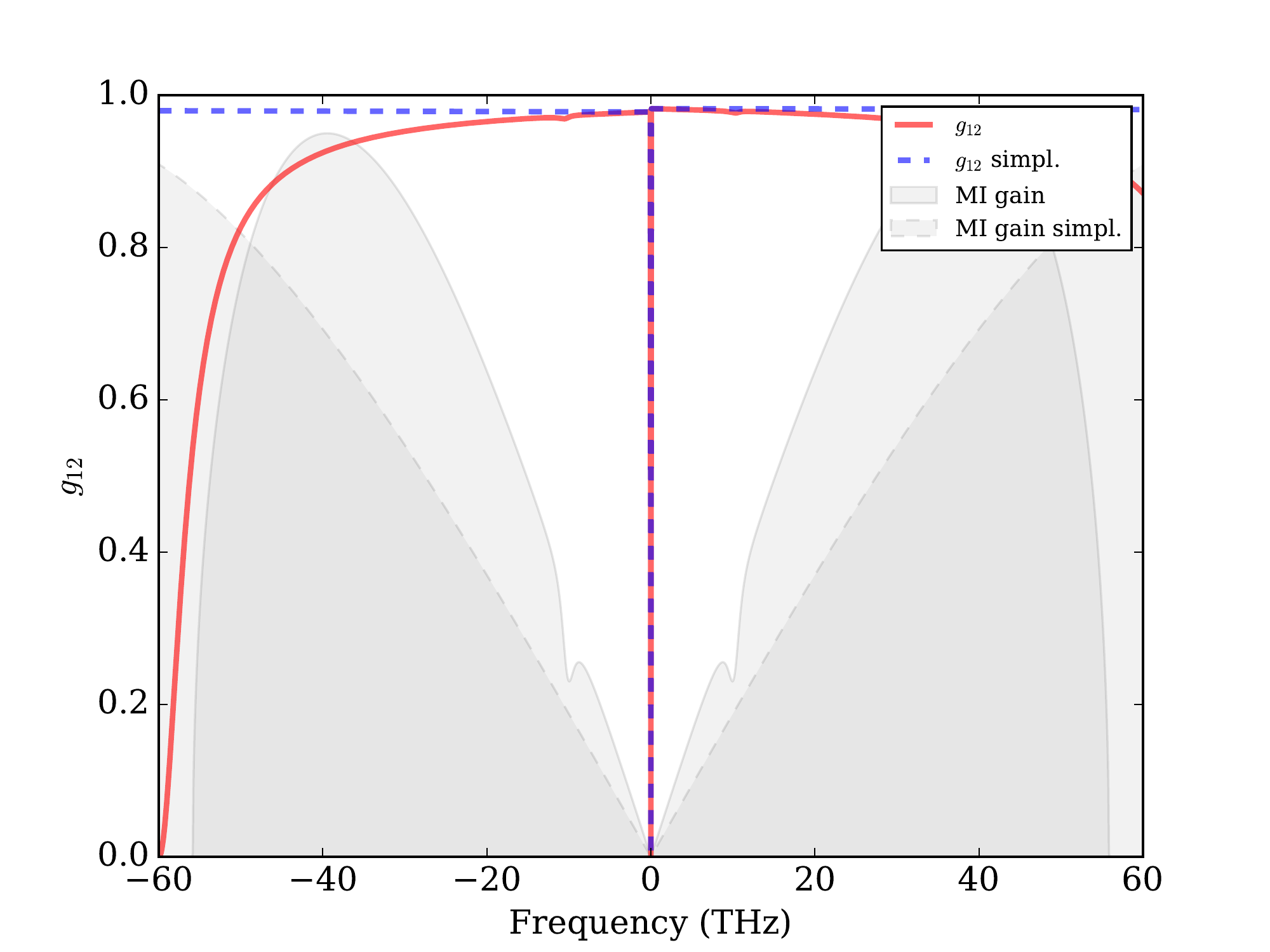}
\caption{Idem Fig. 4 in the mid IR spectral region.}
\label{fig:5}
\end{figure}

In both Fig.~\ref{fig:2} and Fig~\ref{fig:3} a complete model that includes Raman and self-steepening is used, whereas in Fig.~\ref{fig:4} we compare the complete model for SSMF of Figs.~\ref{fig:1}--\ref{fig:2} at distance $2 L_{\mathrm{NL}}$, to the simplified case where neither Raman nor self-steepening, nor higher-order linear dispersion are present. The difference is nearly negligible. To show that this is not always the case, in Fig.~\ref{fig:5} we consider the same comparison as in Fig.~\ref{fig:4}, but for a typical chalcogenide fiber (see, \textit{e.g.}, \cite{Dantanarayana.OptMatExpress.2014,Karim.OptExpress.2015}) in the mid-IR spectral region, where the influence of self-steepening is greatly augmented. Here, as it can be readily observed, not considering a complete model leads to wrong results. 

\begin{figure}
\includegraphics[width=\linewidth]{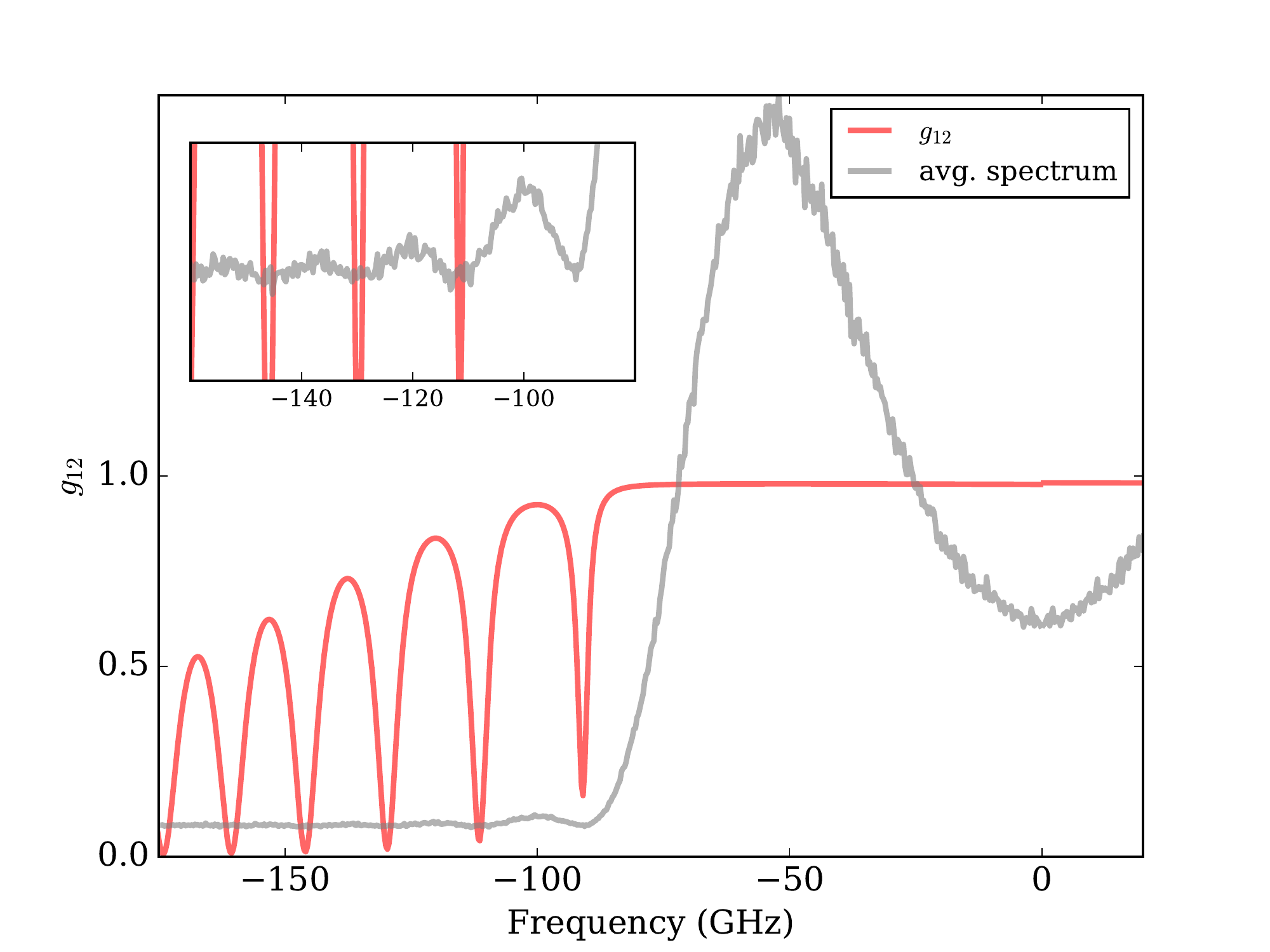}
\caption{$g_{12}$ versus analytic averaged spectra. Zeroes in the coherence function follow ripples in the spectrum.}
\label{fig:6}
\end{figure}

\begin{figure}
\includegraphics[width=\linewidth]{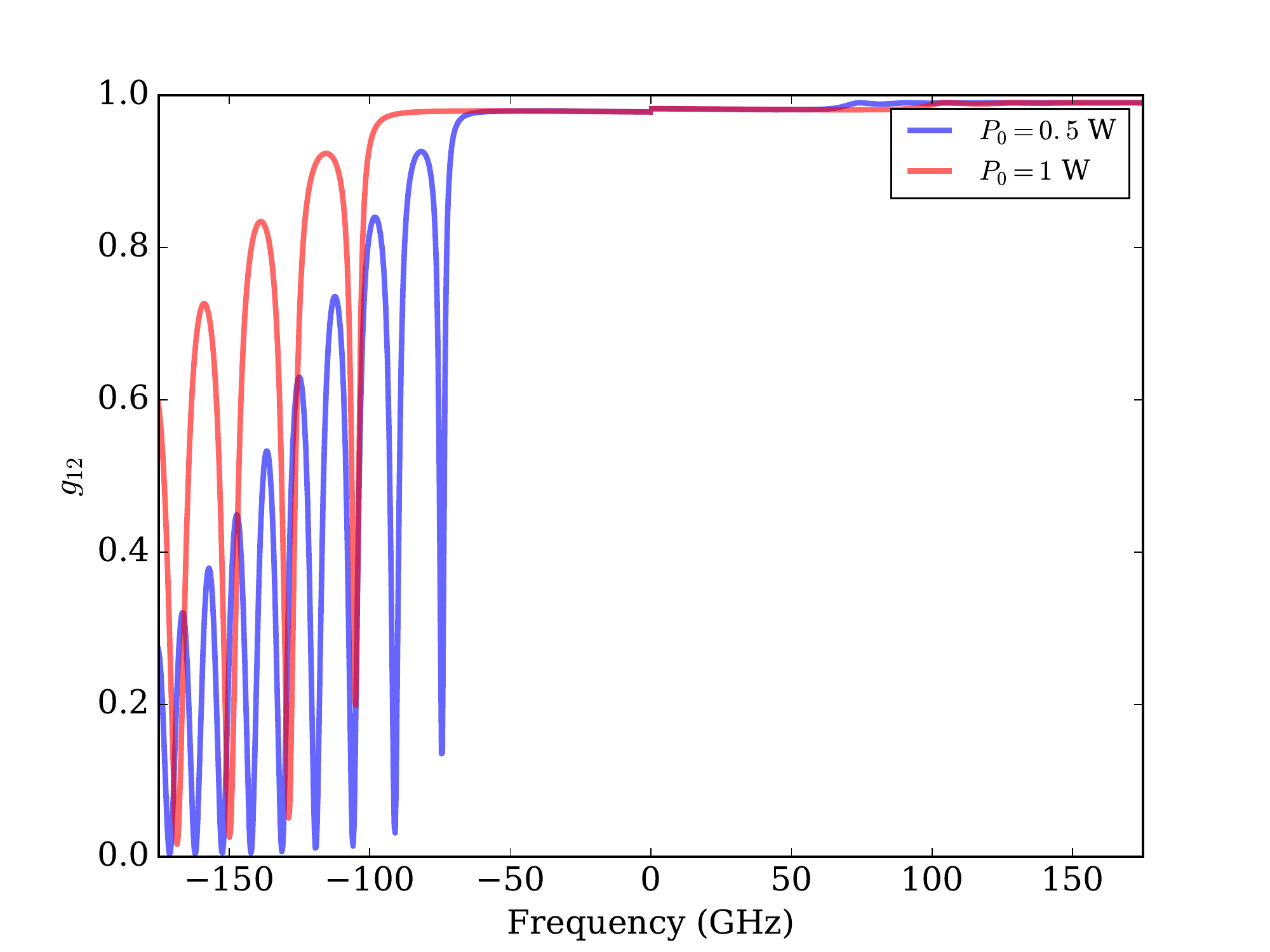}
\caption{$g_{12}$ for different pump powers at $2 L_{\mathrm{NL}}$ (NZ-DSF fiber).}
\label{fig:7}
\end{figure}

It is interesting to point out that, although higher coherence is tightly connected to the region where there is MI gain, the presence of decreasing lobes of considerable coherence beyond the MI cutoff frequency becomes apparent. In Fig.~\ref{fig:6} we show $g_{12}(\Omega)$ against a scaled plot of the averaged spectra of 2000 noise realizations. A strong correlation between $g_{12}$ side lobes and spectral ripples is evident. Also, the MI gain cutoff frequency grows with increasing power. As such, Fig.~\ref{fig:7} shows the enlargement of the coherence bandwidth when increasing the pump power from 500 mW to 1 W.

\begin{figure}
\includegraphics[width=\linewidth]{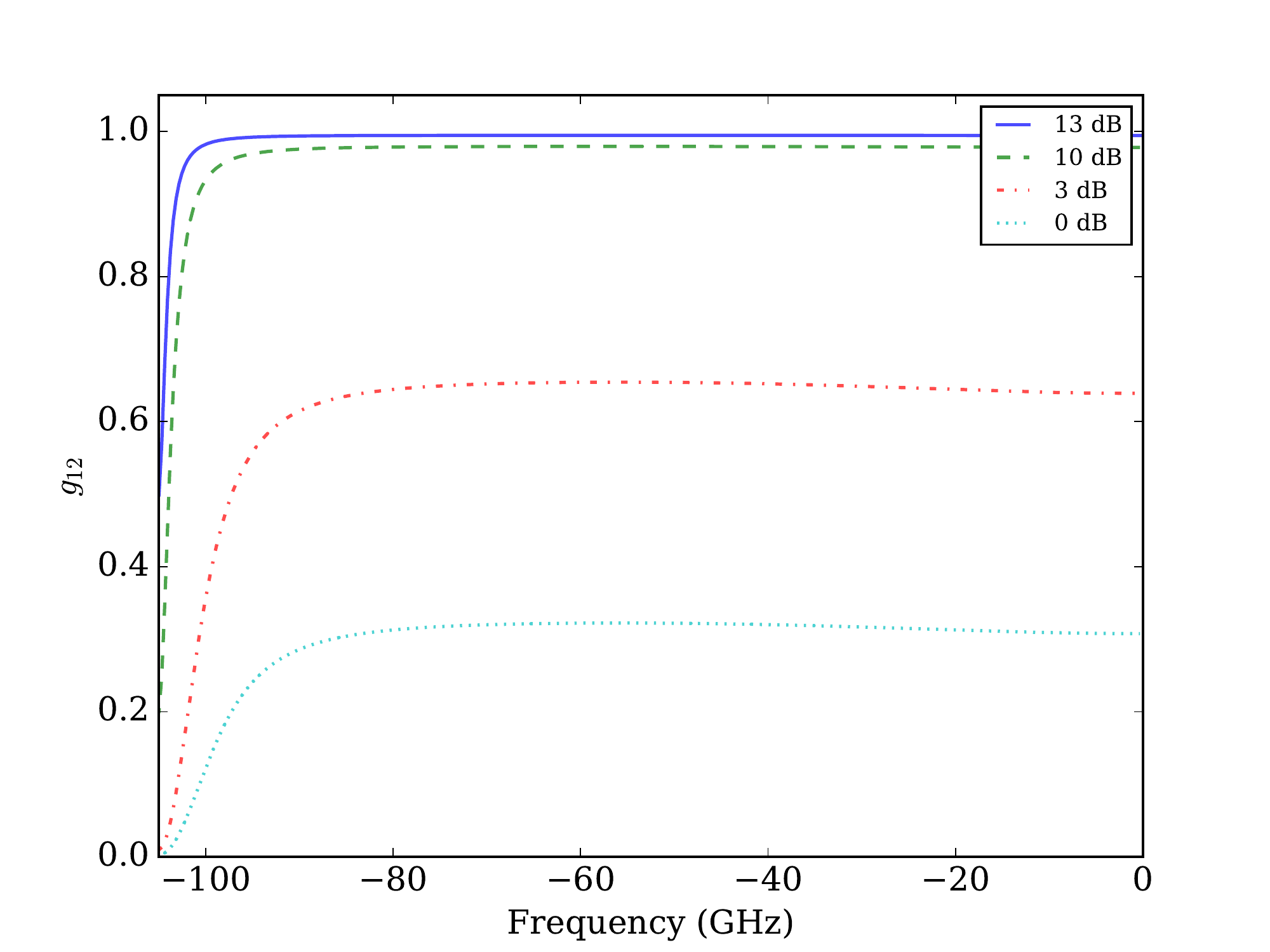}
\caption{$g_{12}$ Stokes-side `coherence bandwidth' for different signal-to-noise ratios at $2 L_{NL}$ (NZ-DSF fiber).}
\label{fig:8}
\end{figure}

Finally, when fixing both pump power (1 W) and distance ($2 L_{\mathrm{NL}}$), varying the seed signal-to-noise power ratio has little effect on the the coherence bandwidth; however, the 3-dB point (as measured from $\Omega=0$ to the point where the maximum $g_{12}$ has decreased to $g_{12}/2$) grows monotonically, as it can be seen in Fig.~\ref{fig:8}.

\section{Conclusions}
\label{sec:conclusions}

In this paper we derived analytical expressions for the coherence in the onset of modulation instability in excellent agreement with numerical simulations. We focused on the case where broadband noise is added to a continuous wave pump (with wavelength lying in the telecommunication near IR region) and investigated the effect of adding a deterministic seed to the CW pump, a case of singular interest as it is commonly encountered in parametric amplification schemes. We obtained results for the dependence of coherence on parameters such as fiber type, pump power, propagated distance, and the signal-to-noise ratio of the seed wavelength. We also investigated the coherence bandwidth, finding that non-zero dispersion shifted fibers yield a greater bandwidth than standard single-mode fibers. Interestingly, in both cases, the coherence bandwidth extends beyond the MI gain cut off frequency.
Finally, We found that inclusion of higher-order linear and nonlinear dispersion leads to very similar results to the simplified model in the near IR range, but is of critical importance when looking at higher wavelengths (mid IR range), where the influence of the self-steepening effect is greatly augmented.
We believe our results to be of significance when analyzing the onset of supercontinuum generation from CW pumps, as they shed light into the coherence of the initial evolution stage.

\bibliographystyle{apsrev4-1}
\bibliography{biblio}

\newpage

\newpage

\end{document}